**Interplay between the glassy transition and granular superconductivity in organic materials**

S. Senoussi,[a][1]  A. Tirbiyine,[2] A. Ramzi,[2] A. Haouam,[3] and F. Pesty[1]

[1] Laboratoire de Physique des solides (CNRS, UMR8502), Université Paris-Sud 11, Bât. 510, 91405 Orsay, France

[2] Laboratoire des Matériaux Supraconducteurs, Université Ibn Zohr, B.P. 8106, Agadir, Morocco

[3] Laboratoire des Couches Minces et Interfaces, Université Mentouri 2500 Constantine, Algeria

**Abstract**

It is known that some (BEDT-TTF)$_2$X layered organic superconductors undergo a glassy transition near 80 K. Our purpose is to exploit quenched disorder to get new insights on both the superconducting state ($T \leq 12$ K) and the glassy transition by studying the superconducting properties as functions of annealing time ($t_a$) and temperature ($T_a$) around 80 K. The main results on the fully deuterated $\kappa$-(BEDT-TTF)$_2$Cu[N(CN)$_2$]Br compound are: 1) The data can be described by a percolation cluster model. 2) At short time scales, the clusters grow with $t_a$ following a power law. 3) At large time scales the clusters grow toward a thermodynamic state following a stretched exponential law $\propto \left(1 - \exp\left(-(t/\tau)^\beta\right)\right)$ with $\beta$ varying from about 0.5 to 1 in our $T_a$ range (65 – 110 K). 4) The relaxation time follows an Arrhenius law $\tau(T) = \tau_0 \exp(U/T)$ with $U \approx 2660$ K and $1/\tau_0 \approx 2 \times 10^{13}$ s$^{-1}$. 5) The asymptotic magnetization fits with a scaling law $\propto (T_a - T_g)^{-n}$ with $T_g \approx 55K$ and $n \approx 3.2$. The results are consistent with a Ising spin-glass-like model.

[a] Corresponding author. E-mail: senoussi@lps.u-psud.fr

It is known [1-10] that the organic conductor $\kappa$-(BEDT-TTF)$_2$Cu[N(CN)$_2$]Br (here, BEDT-TTF is bis(ethylenedithio)tetra-thiafulvalenium while Cu[N(CN)$_2$]Br is a monovalent anion) undergoes a structural transition involving the ethylene groups), also called glassy transition, at a certain temperature (70-80 K) and becomes superconductor below about 12 K. Moreover, in some circumstances this material exhibits a magnetic transition around 10-20 K preceded by the onset of magnetic fluctuations at about 50 K [11,12]. More precisely, at higher temperature the ethylene groups oscillate rapidly between two different conformations. Upon cooling, these thermal fluctuations gradually slow down. Simultaneously, a kind of long-range order among the ethylene groups builds up by choosing one of the two conformations. However, we emphasize that the low-temperature state contains some amount of quenched disorder that strongly depends on the anion, the applied pressure, the cooling conditions and, more importantly, on whether the ethylene groups are formed by hydrogen bonds (denoted H8-Br) or deuterium bonds (denoted D8-Br). This is why the exact nature of the 80 K anomaly and the associated low-temperature superconducting state are still not understood. As a matter of fact, D8-Br exhibits several puzzling superconducting properties. Some of these properties were ascribed to the appearance of a magnetic transition at low temperature (10-20 K), the meaning of which is unclear: Several mechanisms have been proposed including spin-density waves [13], spin canting [12], or the suppression of superconductivity by dispersed magnetic ions associated with the persistent disorder [14]. Also, some authors [7,8] ascribed the decrease of the apparent superconducting volume to an increase in $\lambda$ (the in-plane penetration depth) through mean-free-path effect [7].

However none of these mechanisms is able to explain, even qualitatively, the following experimental facts: 1) The fact that the apparent magnetic critical current density $j_c$ (as deduced from the Bean model) always decreases as the degree of quenched disorder increases. 2) The low-field susceptibility also decreases in the same experimental conditions. To account for such a decrease, one is led to assume that $\lambda$ can reach unphysical values, as high as 0.1 mm [8] (and much more if applied to the present results). 3) Despite these large changes in the apparent $j_c$ and $\lambda$, $T_c$ stays almost constant. 4) As the quenched disorder is increased; one observes: firstly, the appearance of increasing irreversibilities at $H << H_{c1}$ (for instance, $H$ = 0.1 Oe and $T$ = 2 K), secondly, a large broadening of the superconducting transition.

One of the purposes of this work is to prove that all of these features can naturally be explained in the framework of a granular model and the associated weak links connecting the grains.

We have investigated the ac magnetic susceptibility and the dc magnetization of the superconducting phase after performing a series of thermal treatments on a D8-Br single-crystal of dimension ~ 1 × 1 × 0.25 mm$^3$. To this end, we annealed the sample inside the SQUID cryostat at different fixed temperatures $T_a$ ranging from 65 to 110 K. Moreover, in order to carefully monitor the development of ethylene ordering as a function of time at a given $T_a$, we varied the cumulated annealing time $t_a$ from about 30 seconds at 100 K to about $2 \times 10^6$ sec at 65 K. At each $T_a$ the cumulated time $t_a$ was divided into a sequence of intervals (say $\Delta t_w$) ranging from about 20 seconds at 95 K to 12 hours at 65 K. At the end of each interval $\Delta t_w$ the sample was rapidly cooled (20 K/min) to 2K. The ac susceptibility $\chi(T)$ (ac field $h$ = 3 Oe) and the magnetic hysteresis cycle $M_i(H)$ (0 < $H$ < 1000 Oe, typically) were subsequently measured between 2 and 15 K using a Quantum Design SQUID (in both measurements the field was perpendicular to the conducting planes). Then, the sample was again rapidly warmed up (10 K/min) to the same $T_a$, after which the same annealing-cooling-measuring procedure was repeated. Note that the fastest cooling rate was limited by the cooling power of the SQUID and yielded a systematic error in the origin of the annealing time. However, we found this to be negligible below about 85 K.

The initial idea behind this work was to use the extreme sensitivity of the superconducting properties to any kind of disorder to get an insight into this very rich though complicated situation. Qualitatively speaking, we can consider two limits of such disordered structures: the simplest one is the usual punctual defects diluted in a perfect crystal. The other one corresponds to a granular (or cluster) like structure, a typical example of which is provided by high-$T_c$ sintered ceramics such as $YBa_2Cu_3O_{7-\delta}$ [15]. The former kind of defect is expected to have little influence on the low-$H$ ($H << H_{c1}$) low-$T$ ($T << T_c$) magnetic susceptibility (i.e., on the Meissner effect) but would proportionally enhance the critical current density and the associated irreversible magnetization $M_i$ via vortex pinning ($H >> H_{c1}$). It would also reduce the susceptibility transition width near $T_c$ (since pinning delays the entry of vortices into the sample). On the other hand, a cluster (or granular) structure is expected to alter both $\chi$ and $M_i$ (at any $H$ and $T$) and would, in addition, give rise to Josephson-like effects depending on the ratios $\lambda/r$ and $\xi/d$. Following the usual notations $\lambda$ and $\xi$ are the London penetration depth and the superconducting coherence length, while $r$ and $d$ are the average radius of the grains and the grain boundary thickness, respectively. As inferred from the data below, we find that the only way to explain the experimental results is the cluster-like description.

Figure 1 shows both the in-phase ($4\pi\chi$, lower plots) and the normalized out-of-phase ($\chi''$, upper plots) ac susceptibilities as functions of temperature (2-15 K) for $t_a$ ranging from about 5 to 7260 minutes and for $T_a = 69$ K. Taking into account the demagnetizing factor for our parallelepiped geometry, $N \approx 0.65$ [16], we find $-4\pi\chi(1-N) \approx 0.84$ for the longest annealing time. Diamagnetic shielding is still below its maximum possible value, defined by $-4\pi\chi(1-N)=1$. However, the missing 16% are accounted for by the fact that the ratio $r/\lambda$ is not infinite. Indeed, an exact calculation for spherical particles with a ratio of about 20 fully explains the difference. Considering again the $\chi$ curves we observe a large broadening of the superconducting transition, especially for the most rapidly cooled cases. In addition, as illustrated by the imaginary susceptibility $\chi''$ (upper panel), this broadening is accompanied by significant magnetic losses, particularly for short annealing times. Then, since for homogeneous systems we expect negligible hysteresis for measuring fields smaller than the first critical field $H_{c1} \approx 20$ Oe [17], the most plausible explanation for such hysteresis effects would be the presence of weak links between adjacent clusters, with a large distribution of the associated Josephson critical fields [15]. The observed broadening is not necessarily due to inhomogeneities in the temperature transition $T_c$ but can principally be explained by the following mechanisms inherent to any cluster model. Firstly, it is known that when $\lambda(T) \gg r$, $\chi$ varies as $(\lambda/r)^2$ [15,18]. Then, since $r$ is widely distributed and tends to zero with the annealing time, this yields a natural broadening that can be defined by the condition $\lambda(T) > r$. A second source of broadening is connected with the distribution and the temperature dependence of the Josephson critical fields [15,19]. Thirdly, we also expect strong variations in $T_c$ when $r$ becomes comparable with the bulk coherence length $\xi$. However, for strong type II superconductors as is the case here ($\kappa=\lambda/\xi>140$, $\lambda\approx0.7\,\mu m$, Ref. [17]), isolated clusters of such sizes are hard to detect in standard magnetic measurements.

Figure 2 displays $M_i$ vs. $H$ curves obtained at 2 K after following the same annealing procedures as in Fig. 1 ($M_p$ denotes the magnetization at the peak of a cycle.) Here too, we observe a gradual improvement in $M_i$ with $t_a$. In addition, from the $H$-dependence of the $M_i$ curves, we deduce that the pinning strength also increases with $t_a$. The inset to Fig. 2 is an expansion of the low-$H$ cycle for two representative annealing times: $t_a \approx 5$ and 1800 minutes (note the difference of about 18 in vertical scales). Obviously, the low-$H$ behavior of the most annealed state (filled symbols) is perfectly reversible (to within the experimental limit) and linear, while that of the quenched state (open symbols) exhibits strong irreversibilities even at $H$ as low as 1 Oe. For the same physical reasons as for $\chi$, the presence of hysteresis reflects the increasing role of weak links at short annealing times. The increase in the pinning energy with $t_a$ can also be explained by the gradual reduction of the number of such weak links with increasing $t_a$. Finally, all these results confirm that the superconducting clusters and the non superconducting ones are highly entangled.

We now present in Fig. 3 the ac susceptibility $\chi$ together with the peak magnetization $M_p$ as functions of the annealing time at 72, 69 and 65 K. Let us first focus on the 69 and 72 curves from which we draw the following four properties. 1) The relaxation is more rapid for $\chi$ than for $M_p$. 2) Both $\chi$ and $M_p$ tend toward saturation. From now on we shall call these saturation limits $M_s$ and $\chi_s$. 3) $\chi_s$ is nearly the same for 72 and 69 K while $M_s$ changes by a factor of almost two. 4) More fundamentally, these saturation values are found to be reproducible and independent of the previous thermal history (see inset to Fig.3 ). This suggests that they correspond to an equilibrium thermodynamic state.

The examination of the 69 K curves reveals that both $\chi$ and $M_p$ follow two consecutive regimes as functions of $t_a$. Firstly, an initial time regime described by a power law. Such a regime is preponderant below 66 K, but becomes practically inaccessible above $T_a = 72$ K (because of the finite cooling-heating rates). It is spectacularly illustrated by the 65 K curves, which also reveal a considerable slowing down of the relaxation rates below about 69 K. Secondly, a long time regime described by a stretched exponential law $\propto \left(1 - \exp\left(-(t/\tau)^\beta\right)\right)$ (represented by solid lines on the $M_p$ curves). It corresponds to the saturation limit just discussed. This regime is dominant above 72 K but out of reach experimentally below 63-64 K (as it needs a $t_a$ of several years). Such a two regime behavior is common in many fields of physics and is known as "growth and coalescence". In spin glasses the short time regime is associated with a "fined-grained structure" while the long time one, called $\alpha$ regime, defines the single exponential and the stretched exponential regimes [20,21]. Considering the long time regime we find $\beta \approx 1$ for $T \geq 69$ K and $\beta \approx 0.5 \pm 0.1$ at 66 K (note that the fitting functions are shown as solid lines for $M_p$ and dashed lines for $\chi$).

Figure 4 shows the variation against $T_a$ of both $M_s$ and $\chi_S$, calculated by extrapolating the fitting curves as explained before. We see again that the two quantities vary quite differently with decreasing $T_a$: below 70 K, $\chi_S$ tends toward a plateau whereas $M_s$ increases more and more rapidly. This reflects an anomalous increase in the cluster size as $T_a$ decreases and confirms further the cluster description. Indeed, the length scales are very different for the two quantities: $\chi_S$ is expected to saturate for $r \gg \lambda \approx 0.7\,\mu\text{m}$ while $M_s$ would continue increasing up to $r \approx R \approx 500\,\mu\text{m}$ (Bean model, see below). Moreover, we are able to fit $M_s$ with a scaling law of the form $M_s = M_0 \left(T/T_g - 1\right)^{-n}$ (continuous line in Fig. 4) with $T_g \approx 55$ K, $n \approx 3.2$ and $M_0 \approx 0.16$ emu cm$^{-3}$. The inset shows the long time relaxation $\tau$ (associated with $M_i$) as a function of $T_a$ in a semi-logarithmic scale. The data fit remarkably well with an Arrhenius function $\tau(T) = \tau_0 \exp(U/T)$ with an activation energy $U$ of about 2660 K and an attempt frequency $1/\tau_0$ of about $2 \times 10^{13}$ s$^{-1}$. Our $U$ value is in excellent agreement with NMR data [11] at 200-350 K and to a lesser extent with resistivity measurements [4]. However, our results differ qualitatively and quantitatively from other transport data [2]. Combining the NMR frequencies reported in Ref. [11] and our time scale shows that the $\exp(U/T)$ law is obeyed over 15 orders of magnitude. By contrast our attempt frequency is two orders of magnitudes smaller than that deduced from NMR. However, we have experimental evidence that this difference is a mere isotopic effect.

In most publications on the subject, the interpretation of susceptibility measurements are based on the classical Meissner formula $4\pi\chi = -v_s$ (where $v_s$ is the fractional superconducting volume). In the same way, the irreversible magnetization $M_i$ and the associated $J_c$ are assumed to be related by the Bean model $M_i = v_s \times R \times J_c / 30$, where $R$ is the macroscopic radius of the sample. These formulae implicitly neglect any granular and any weak link effects. However, if one tries to interpret our data in such a picture, one is led to unphysical values for both $\lambda$ and $J_c$ (i.e., $\lambda$ could be as high as 1 cm and $J_c$ as low as 100 A/cm$^2$). In addition, such a model can explain neither the low-$H$ irreversibility nor the transition broadening. Actually, our data are consistent with a granular interpretation where the macroscopic radius $R$ must be substituted by some average cluster radius $r$. Moreover, the topology of the clusters can be described by a percolation model [20] and/or a Ising Spin Glass-like model [21] according to which the cluster size grows as $\langle r^2 \rangle \approx C \left(1 - \exp\left(-(t/\tau)^\beta\right)\right)$. Here $\tau$ and $\beta$ depend on the dimensionality of the material and the treatment temperature: The model predicts that as $T$ is lowered, $\beta$ would vary from $\leq 1$ at $T \geq T_p$ to 1/3 at $T << T_p$ (stretched exponential) where $T_p > T_g$ is the percolation threshold temperature. Our data are consistent with $\beta \approx 1$ at $T \geq 69$ K and $\beta = 0.5 \pm 0.1$ at 66 K. In fact, we observe a dramatic slowing down of $\tau$ for $T \leq 66$ K suggesting that $T_p$ is close to this temperature. Moreover, as predicted by the spin glass model [21] we observe that this slowing down is accompanied by the onset of a two relaxation regime. Accepting again the analogy with spin glasses, our data suggest the presence of two critical points $T_p$ and $T_g$, corresponding to a percolation threshold and to a true thermodynamic transition [22], respectively. At this point, it is worth noting that the deduced $T_g$, $\approx 55$ K, lies in the temperature region where ferromagnetic fluctuations take place, while the resistivity behavior changes from semi-conducting to metallic. Using a granular analysis [15,18,19], we find that at equilibrium $r$ varies from about 0.1 μm at $T_a \approx 100$ K to 10-30 μm at 66 K.

Moreover, the present results prove clearly that the superconducting properties are determined by the quenched disorder created while crossing the glass transition and not by a possible onset of magnetic effects below this transition. The analogy with spin-glasses is straightforward: like Ising spins, ethylene molecules have only two allowed states. In addition, the canonical RKKW interactions can be simulated by a random distribution of incommensurate CDWs [23,24].

At this point, it may seem surprising that we did not refer to the Mott-Hubbard model [25,26], which is known to describe some fundamental properties of organic conductors. This is particularly true for the influence of pressure on various physical parameters [25,27,28]. However, the situation seems to be much more complicated concerning the kind of measurements reported here, even though these measurements are probably connected somehow to the Mott-Hubbard physics. Indeed, in its usual theoretical formulation this model is defined by only two parameters: the inter-site hopping energy and the intra-site Coulomb interaction between electrons. Therefore, in its usual form this model cannot account for ethylene ordering and thermal fluctuations that involve the whole mass of ethylene groups. This is why many experimental data differ considerably when hydrogen is substituted by deuterium in BEDT-TTF compounds. In addition, the classical Mott-Hubbard transition would be first order and would have no order parameter. On the contrary, our experimental results and analysis suggest two ordering parameters: A spatial one, connected with the average size of the clusters, and a temporal one, connected with spin-glass-like relaxation (Edwards-Anderson order parameter [22].

Clearly, the relationship between these spin-glass-like effects and the Mott-Hubbard transition is beyond the scope of this paper. Obviously, this question needs more theoretical and experimental developments.

**Acknowledgements**

The authors wish to thank C. Pasquier for helpful discussions, and C. Mézière and P. Batail for providing the samples. We also thank CNRS (France) and CNRST (Morocco) for financial support via PICS 522.

**Figure captions**

FIG. 1. In-phase ($\chi$, lower part) and out-of-phase ($\chi''$ in relative units, upper part) ac susceptibilities vs. temperature ($h = 3$ Oe), plotted after annealing then quenching (from $T_a = 69$ K) as described by the labels in the figure. For the sake of clarity, only half of the registered curves is displayed. The solid lines are guides to the eyes.

FIG. 2. Magnetization $M_i$ vs. $H$, measured at 2 K, after the same heat treatment (and same symbols) as in Fig. 1. The peak magnetization is denoted $M_p$. The solid lines are guides to the eyes. The inset is an expansion of $M_i$ near $H = 0$ for $t_a \approx 5$ (left scale) and $t_a = 7260$ min (right). Note the difference of 18 in the scales.

FIG. 3. Peak magnetization $M_p$ (filled symbols, right scale) and susceptibility $\chi$ (open symbols, left scale) at 2 K vs. annealing time at $T_a = 65$ (triangles), 69 (squares) and 72 K (circles). Note that above 69 K both quantities tend to saturate but more rapidly for $\chi$. The inset shows that the relaxed state does not depend on whether this state is reached from below or above its saturation value. The solid lines represent fitting curves (see text). The asymptotic values deduced from fitting are called $M_s$ and $\chi_s$.

FIG. 4. Saturation values $M_s$ and $\chi_s$ as functions of annealing temperature $T_a$. As illustrated by the solid line, $M_s$ follows a scaling law (see text). The inset is a semi logarithmic plot of the relaxation time as a function of $1/T_a$: the fitting function (solid line) is a stretched exponential with $\beta \approx 1$ for $T_a > 69$ K and $\beta \approx 0.5 \pm 0.1$ at 66 K (see text).

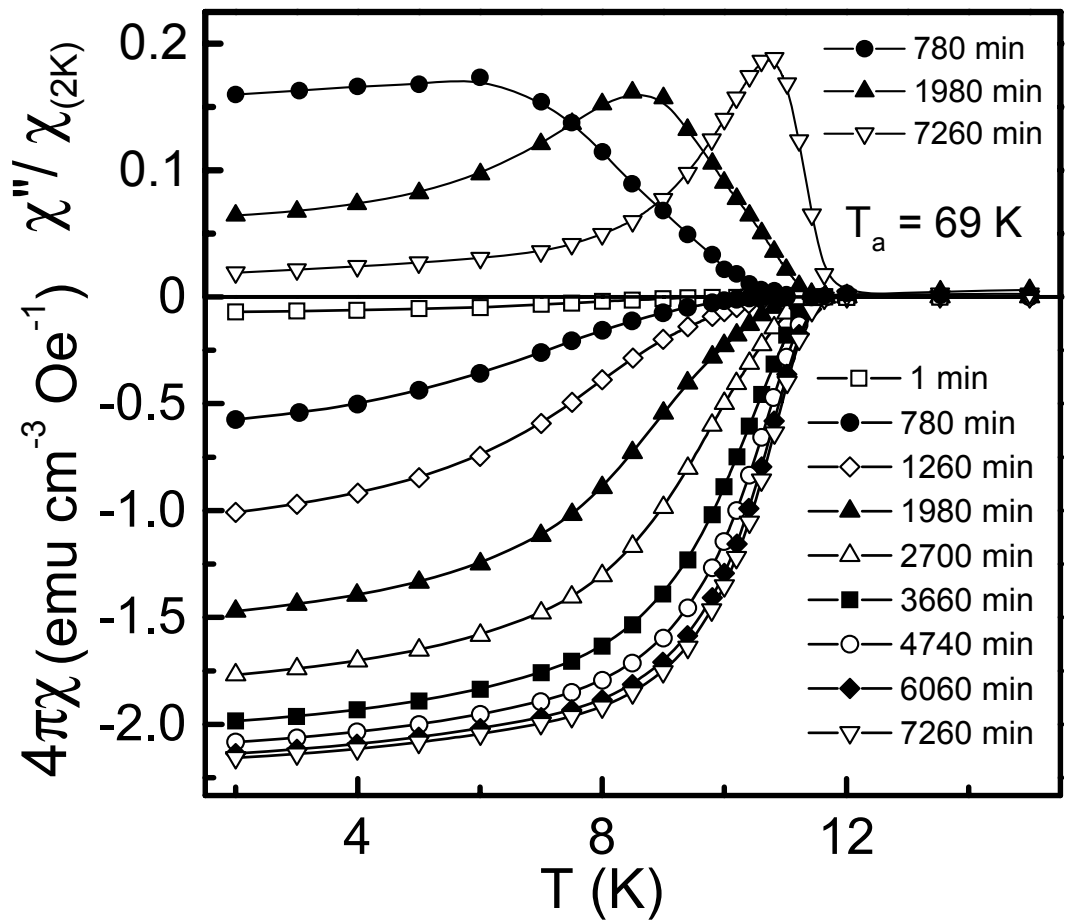

Figure 1

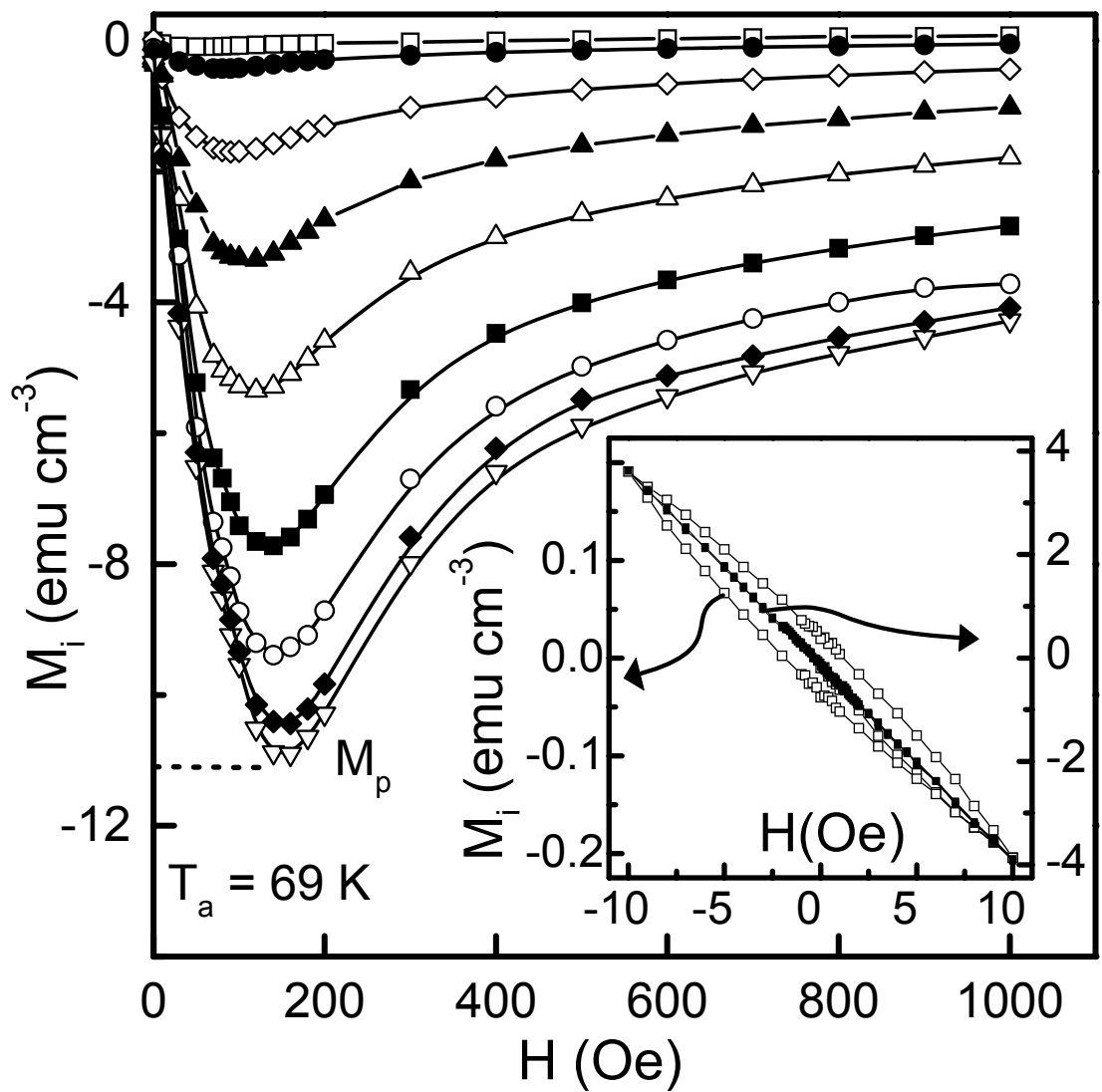

Figure 2

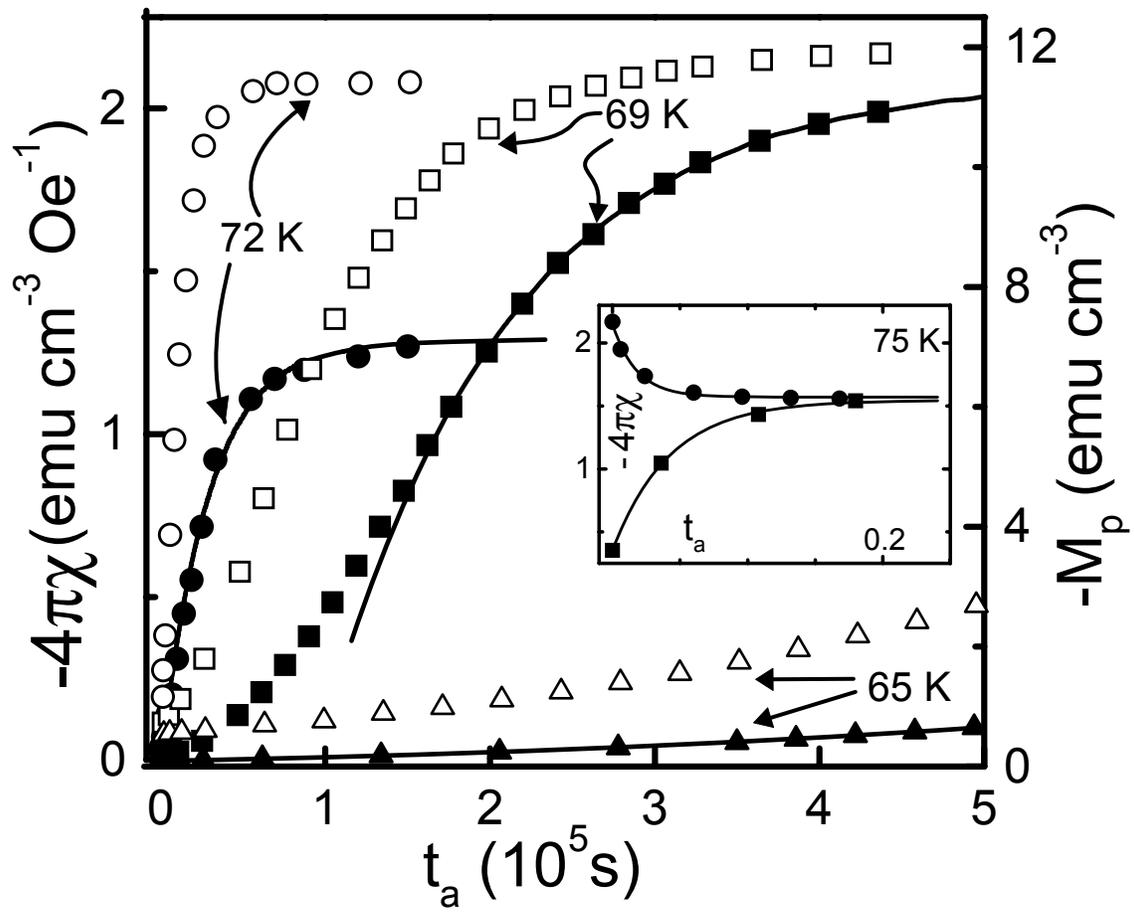

Figure 3

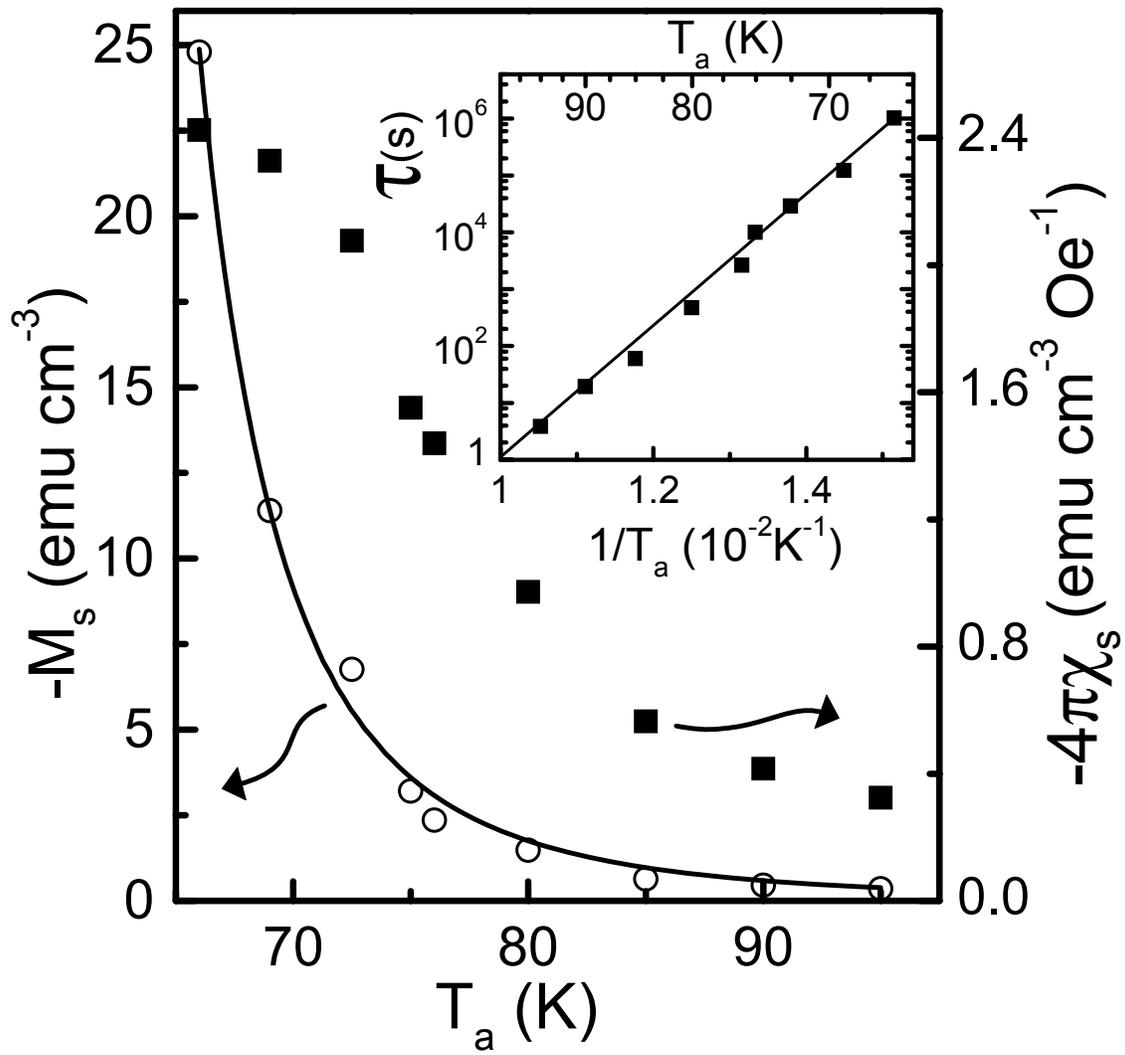

Figure 4